\documentclass[useAMS,usenatbib,usegraphicx]{mn2e}

\usepackage{graphicx,xcolor,textpos}
\usepackage{graphicx,amssymb}
\usepackage{enumitem, amsmath}
\usepackage{wrapfig,caption, url}
\usepackage{subcaption, float, gensymb, soul}
\soulregister\citet7
\soulregister\citep7

\title[Rms-flux relation in CVs]{The rms-flux relation in accreting white dwarfs: \\ another nova-like variable and the first dwarf nova}
\author[M. Van de Sande, S. Scaringi, C. Knigge]
{M. Van de Sande$^{1}$\thanks{E-mail: marie.vandesande@ster.kuleuven.be}, 
S. Scaringi $^{2,1}$,
C. Knigge$^{3}$ \\
$^{1}$ Instituut voor Sterrenkunde, KU Leuven, Celestijnenlaan 200D, B-3001 Heverlee, Belgium \\
$^{2}$ Max-Planck-Institute f\"ur Extraterrestrische Physik, Postfach 1312, D-85748 Garching, Germany \\
$^{3}$ Department of Physics and Astronomy, University of Southampton, Highfield, Southampton, SO17 1BJ, UK}
\begin{document}

\date{Accepted 2015 January 20.  Received 2015 January 20; in original form 2014 December 5}

\pagerange{\pageref{firstpage}--\pageref{lastpage}} \pubyear{2015}

\maketitle

\label{firstpage}

\begin{abstract}
We report on the detection of the linear rms-flux relation in two accreting white dwarf binary systems: V1504 Cyg and KIC 8751494. The rms-flux relation relates the absolute root-mean-square (rms) variability of the light curve to its mean flux. The light curves analysed were obtained with the \emph{Kepler} satellite at a 58.8 s cadence. 
The rms-flux relation was previously detected in only one other cataclysmic variable, MV Lyr. 
This result reenforces the ubiquity of the linear rms-flux relation as a characteristic property of accretion-induced variability, since it has been observed in several black hole binaries, neutron star binaries and active galactic nuclei.
Moreover, its detection in V1504 Cyg is the first time the rms-flux relation has been detected in a dwarf nova-type CV during quiescence.
This result, together with previous studies, hence points towards a common physical origin of accretion-induced variability, independent of the size, mass, or type of the central accreting compact object.

\end{abstract}

\begin{keywords}
accretion discs, binaries: close, stars: novae, cataclysmic variables, stars: dwarf novae, stars: individual: V344 Lyr, V363 Lyr,  V447 Lyr, V516 Lyr, V585 Lyr, V1504 Cyg, KIC 8751494 
\end{keywords}

%%%%%%%%%%%%%%%%%%%%%%%%%%%%%%%%%%%%%%%%%%%%%%%%%%%%%%%%%%%%%%%%%%%%%%%%%%%%%%
\section{Introduction}
%%%%%%%%%%%%%%%%%%%%%%%%%%%%%%%%%%%%%%%%%%%%%%%%%%%%%%%%%%%%%%%%%%%%%%%%%%%%%%

\begin{table*}
\centering
\begin{minipage}{160mm}
\centering
\caption{All known CVs in the \emph{Kepler} field-of-view for which \emph{Kepler} short cadence data is available. The length of their total analysed data sets and the total amount of steady data are given. 
MV Lyr has already been analysed by \citet{Scaringi2012}, all available data is listed here.
{Systems for which the number of steady states is not given did not exhibit any outbursts in the \emph{Kepler} data.}
}
\label{table:allCVs}
\begin{tabular}{l c l c c c c l c}        
	\hline                 
	Name 		& KIC  		& Type  	 	& \emph{Kepler}	& Total		& Steady	& \# Steady & \emph{Kepler}	& Rms-flux	\\    
				& number	&				& magnitude		& data (d)	& data (d)	& states	& quarters		& detected?	\\
	\hline                     	
	BOKS 45906	& 9778689	& DN 			& 19.046		& 434.0		& 319.4		& 6 		& 6-8, 11, 15	& No		\\
	V344 Lyr	& 7659570	& DN - SU UMa	& 14.881		& 1343		& 111.1		& 37		& 2-17			& No		\\
	V363 Lyr	& 7431243	& DN - SU UMa	& 16.672		& 5.210		& 5.210		& -			& 16			& No		\\
	V447 Lyr	& 8415928	& DN - U Gem	& 18.430		& 608.8		& 516.5		& 16		& 8-11, 13-17	& No		\\
	V516 Lyr	& 2436450	& DN			& 18.798		& 163.0		& 88.03		& 11		& 8-9			& No 		\\
	V585 Lyr	& 5523157	& DN - SU UMa	& {19.489}	 	& 191.1		& 191.1		& -			& 9, 14			& No 		\\
	V1504 Cyg	& 7446357	& DN - SU UMa	& 15.805 		& 1316		& 549.4		& 121		& 2-16			& Yes 		\\
				& 8751494	& NL			& 16.269		& 226.5		& 226.5		& -			& 2-3, 5, 15	& Yes 		\\
	{MV Lyr}	& {8153411} & {NL}			& {12.640}		& {1342.6}	& {1342.6}	& {-} 		& {2-17}		& {Yes}		\\
\hline
\end{tabular}
\end{minipage}
\end{table*}

Compact interacting binaries (CBs) are binary systems consisting of a low-mass red dwarf secondary and a compact object, such as a white dwarf (WD), neutron star (NS) or black hole (BH). 
An accretion disc surrounding the compact object is formed due to Roche-lobe overflow of the secondary.
As angular momentum is transported outwards within the disc, material spirals towards the compact object. The accretion disc gives rise to aperiodic broad-band variability, also known as flickering, in all types of CBs \citep{vanderKlis1995, Warner2003}.
The time-scale and energy range at which aperiodic variability can be observed depends on the central compact object. In X-ray binaries (XRBs, accreting NS or BH) the aperiodic variability is mainly visible in the X-ray domain over time-scales ranging from milliseconds to minutes {\citep{vanderKlis1995}}. 
In active galactic nuclei (AGN, accreting supermassive BH) it is mainly visible in the optical and UV domain over time-scales ranging from days to years \citep{Uttley2001, Uttley2005}. Part of the energy is also emitted in the X-ray domain, where aperiodic variability has been previously studied in these systems. 
In cataclysmic variables (CVs, accreting WD) it is visible in the optical and UV domain over time-scales ranging from seconds to hours \citep{Scaringi2012, Scaringi2013a}.
The differences in energy and time-scale are due to the difference in size of their central compact objects and of the inner discs of their accretion discs. 
In XRBs, the inner rim of the disc is located typically at a few km from the central object. Because of the small size of the inner disc, the aperiodic variability emitted by the disc has a larger high-frequency component. Material is also able to reach down deeper into the gravitational potential of the central object, causing it to emit mainly in the X-ray domain.
In AGN, the location of the inner rim is dependent on the event horizon, typically a few AU from the central object. The characteristic time-scales of the emitted aperiodic variability are hence much longer.
In CVs the central object is larger than that of XRBs, such that the inner rim of the disc is located at a few 1000 km. The characteristic time-scales of the emitted aperiodic variability are therefore somewhat smaller than those of AGN, but larger than those of XRBs, while the disc emits mainly in the optical and UV domain.

Aperiodic variability is characterised by several properties, among which the linear rms-flux relation.
This relation relates the absolute root-mean-square (rms) variability of the light curve to its mean flux: the stronger the variability, the brighter the system.
The rms-flux relation implies that short and long variability time-scales are coupled multiplicatively \citep{Uttley2005}. 
Currently, aperiodic variability is modelled best by the fluctuating accretion model \citep{Lyubarskii1997, Arevalo2006, Scaringi2014}. 
In this model, the bulk of the emission of the disc is produced at its inner boundary layer, while aperiodic variability originates throughout the disc and modulates the central emission.
A key feature of the fluctuating accretion model is the multiplication of the various variability components generated throughout the disc, naturally resulting in the coupling of short and long time-scales.
The rms-flux relation rules out shot-noise models, in which the different time-scales within the variability are combined additively.

{Aperiodic variability and its properties have been studied extensively in XRBs and AGN. 
This is mainly due to the availability of X-ray data suitable for its study obtained by the NASA Rossi X-Ray Timing Explorer (\emph{RXTE)} mission \citep{vanderKlis1995}. 
The linear rms-flux relation has therefore been detected in several XRBs and AGN. An exception has been found in a BL Lacertae blazar, which hinted at a divergence from a linear relation at high fluxes \citep{Edelson2013}.}

Due to the longer variability time-scales in CVs, long and uninterrupted optical light curves are needed. It was not until the launch of NASA's \emph{Kepler} satellite that these were available for the study of aperiodic variability in CVs.
So far, \citet{Scaringi2012} have discovered the rms-flux relation in the nova-like variable MV Lyr. 
Nova-like variables are considered to be analogous to high-soft state XRBs \citep{Kording2008}. The rms-flux relation is not yet detected in a low-hard state analogue. 

Here we want to determine whether the rms-flux relation is a ubiquitous feature amongst accreting WDs, as is the case for XRBs and AGN \citep{Heil2012}. In section 2 we will discuss the use of \emph{Kepler} data and the data analysis method. The results are presented in section 3. Our discussion and implications are presented in sections 4 and 5.

%%%%%%%%%%%%%%%%%%%%%%%%%%%%%%%%%%%%%%%%%%%%%%%%%%%%%%%%%%%%%%%%%%%%%%%%%%%%%%
\section{Observations and data analysis}
%%%%%%%%%%%%%%%%%%%%%%%%%%%%%%%%%%%%%%%%%%%%%%%%%%%%%%%%%%%%%%%%%%%%%%%%%%%%%%

\begin{figure*}
	\center
	\begin{subfigure}[t]{0.95\textwidth}
	\centering
	\includegraphics[width=\columnwidth]{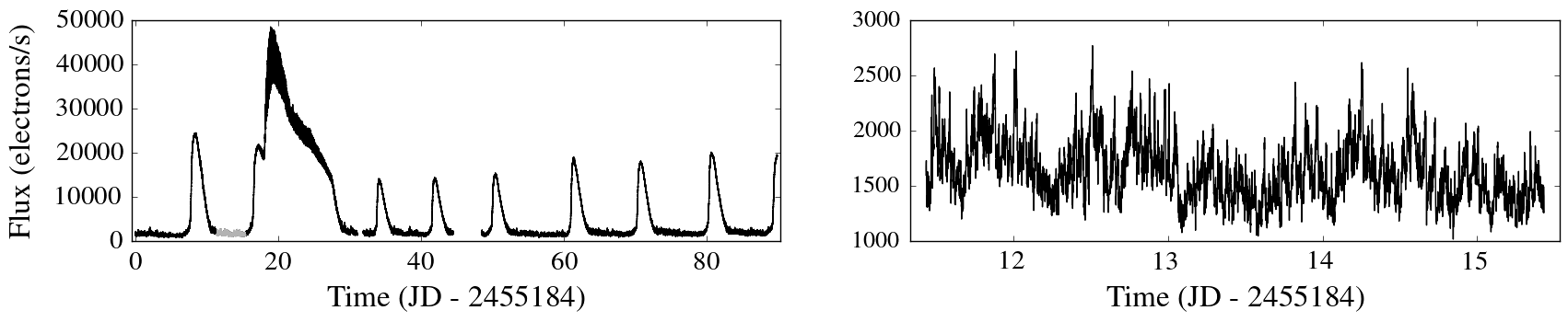}
	\captionsetup{justification=centering}
	\caption{\emph{Kepler} Q5 observations of V1504 Cyg}
	\label{subfig:Data-V1504Cyg-LC}
	\end{subfigure}
	
	\vspace{4mm}

	\begin{subfigure}[t]{0.95\textwidth}
	\centering
	\includegraphics[width=\columnwidth]{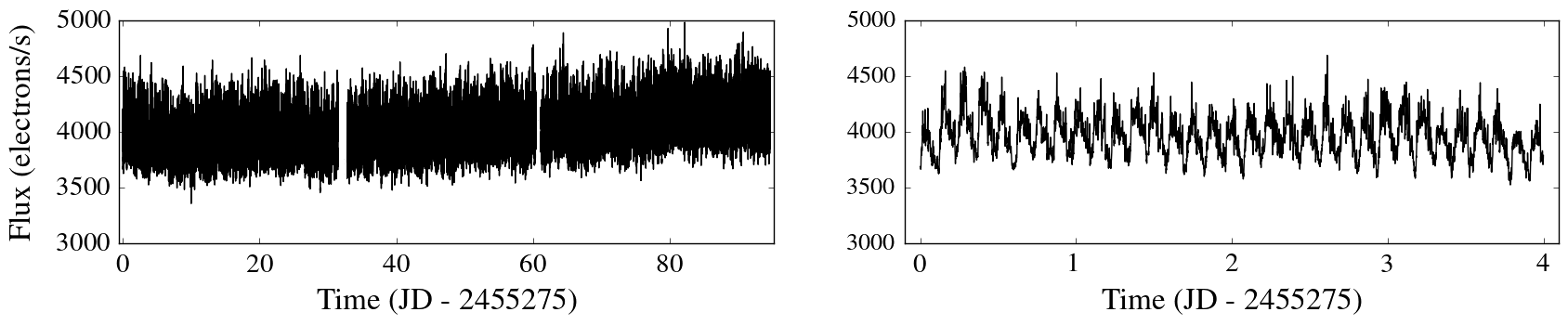}
	\captionsetup{justification=centering}
	\caption{\emph{Kepler} Q5 observations of KIC 8751494}
	\label{subfig:Data-KIC8751494-LC}
	\end{subfigure}	
	\caption{Illustration of the \emph{Kepler} light curves of V1504 Cyg and KIC 8751494, the measurements during \emph{Kepler} Q5 are shown in the left panels.
	The selection of quiescent intervals in DNe is illustrated for V1504 Cyg: in the left panel of figure \ref{subfig:Data-V1504Cyg-LC}, a steady interval has been selected in gray. This is shown in more detail in its right panel.
	For KIC 8751494, a four day-segment is shown in more detail in the right panel of figure \ref{subfig:Data-KIC8751494-LC}, where the orbital period can also be seen. Aperiodic variability is clearly visible in both objects.
	} 
	\label{fig:Data-LC}
\end{figure*}

%-----------------------------------------------------------------------------
\subsection{\emph{Kepler} mission and object selection}
%-----------------------------------------------------------------------------

The NASA \emph{Kepler} mission was successfully launched on March 7 2009. 
It continuously pointed towards the same 116 square degree field-of-view (FOV) during the entire mission, and obtained light curves with a 58.8 s cadence (short cadence, SC) or with a 30 min cadence (long cadence, LC) during every quarter (Q) of observations \citep{Borucki2010}. 
In May 2013, \emph{Kepler} entered a prolonged safe mode due to a second reaction wheel failure. 
The satellite is fortunately not lost: the \emph{K2} mission became operational in June 2014 and will continue to gather data of similar quality similar to the original \emph{Kepler} mission. {Unlike the original mission, \emph{K2} will cover new areas of the sky and will hence not obtain the original extensive, uninterrupted temporal coverage of a single FOV \citep{Howell2014}.}

Several CVs can be found in the \emph{Kepler} input catalogue (KIC). 
\citet{Still2010} list all objects found in both the input catalogue and the CV catalogue made by \citet{Downes2001}. Additions to the list of \citet{Still2010} were made by \citet{Williams2010}, \citet{Feldmeier2011}, \citet{Ramsay2012}, \citet{Scaringi2013} and \citet{Ramsay2014}. \citet{Howell2013} checked spectroscopically if these objects were indeed CVs.
All CVs in the \emph{Kepler} FOV for which SC light curves were available are listed in table \ref{table:allCVs}.
Only the {simple} aperture photometry (SAP) light curves are considered, as the pre-search data conditioned light curves are optimised for planet transit searches{, which removes long term variability and could affect temporal analyses.}
The nine CVs listed in table \ref{table:allCVs} are all known CVs in the \emph{Kepler} FOV for which SC light curves are available, and consist both dwarf novae and nova-like variables (see below).
{MV Lyr has already been analysed \citep{Scaringi2012, Scaringi2013a, Scaringi2013b}, but is included in the table for completeness.}

%-----------------------------------------------------------------------------
\subsection{Data selection}
%-----------------------------------------------------------------------------

Our sample consists of dwarf novae (DNe) and nova-like variables (NLs). 
DNe experience regular outbursts which are caused by instabilities within the accretion disc. Within this class, one can differentiate between SU UMa and U Gem stars. The former experiences superoutbursts as well as regular DN outbursts, while the latter does not.
It is important to remove those large amplitude outbursts from the aperiodic variability analysis, as they will produce overestimates of the rms.
This has been done by visually inspecting all available data and by using conservative ranges when selecting the intervals, in order to ensure that the systems are fully in their quiescent states. 
For NLs, which do not experience such events, every quarter can be analysed as a whole.
Quiescent states of DNe and observations of NLs will both be called `steady' states or quarters.
Obvious outliers, which can be caused by systematic error sources from the telescope and spacecraft \citep{Fraquelli2012}, are removed from all steady data sets. 
{This has been done by visually analysing the flux distribution of each light curve, plotted on a semi-logarithmic scale. Using this scaling, any discrepancies to the flux distribution can be spotted and removed from the light curve.}
In the left panels of figure \ref{fig:Data-LC}, we show the light curves observed for V1504 Cyg and KIC 8751494 during \emph{Kepler} Q5. Various DN outbursts and one superoutburst are clearly visible in the light curve of the DN- SU UMa V1504 Cyg (figure \ref{subfig:Data-V1504Cyg-LC}). The selection of a quiescent interval is shown as well in gray, which is shown in detail its right panel.
As is clear from figure \ref{subfig:Data-KIC8751494-LC}, the NL KIC 8751494 does not show any outbursts. Aperiodic variability is clearly visible in both objects (right panels of figure \ref{fig:Data-LC}).

{Temporal gaps in the light curves are created by removing outbursts or are due to gaps in the \emph{Kepler} observations itself. As our data analysis method (see section {\ref{subsect:dataanalysis}}) is insensitive to gaps in the data, we did not join different segments separated by a gap.}
The steady intervals of DNe can be combined according to their quarter of observation. 
{This also creates temporal gaps in the light curve, which we again keep as part of the data set. Small offsets due to long terms drifts in the light curve were noted, but were small enough to have a negligible impact on our method.}
However, we cannot combine all their steady intervals over several quarters or combine all quarters of NLs, since the mean flux level is shifted when entering a new quarter. 
{These shifts are due to the quarterly 90$\degree$ rolls the satellite had to perform in order to have its solar panels face the Sun. After every roll the object is placed on a different CCD camera, which causes the optimal aperture to change. It is the change in aperture that causes the inter-quarter flux offsets \citep{Jenkins2010}.
We have made no attempt to correct for these inter-quarter flux offsets as we only analyse quarters for NLs and single steady intervals or steady intervals combined according to their quarter of observation for DNe.
It would be interesting to do so in the future.}

%-----------------------------------------------------------------------------
\subsection{Data analysis methods} 		\label{subsect:dataanalysis}
%-----------------------------------------------------------------------------

The data analysis method is applied to every steady data set of each object. The aim is to determine whether the rms-flux relation is present in the steady data sets, and if so, to fit a linear model to it. In order to reduce the error on the intercept of the fitted linear relation, we first subtract the mean flux level from the steady data set.
This does not affect the detection of the rms-flux relation, as we are simply subtracting a constant from the entire data set under consideration.
The data analysis method can be split into two main parts: an appropriate binning of the data set followed by fitting a linear model to the binned results.
We first bin the light curve over time. This is done by dividing the light curve into several segments, each containing the same number of measurements. As every measurement is taken at a 58.8 s cadence, the number of measurements in each bin corresponds to the time-scale of aperiodic variability under investigation.
The number of bins after the first binning is hence dependent on the chosen time scale.
The mean flux value of the measurements and their variance is calculated for each bin.
Within each bin, the expected Poisson noise level is subtracted from the measured variance in order to obtain an indicative measure of the intrinsic variance of the data. The expected Poisson noise level has been calculated for every sampled time scale and hence bin size.
It is known however that \emph{Kepler} light curves suffer from additional noise components, and thus the Poisson noise estimate used is a conservative one. The mean flux - intrinsic variance results are averaged over by rebinning them into ten bins. 
All bins are equally weighted.
We have checked that the number of bins adopted does not significantly affect the final results, and have chosen for ten bins in accordance with \citet{Scaringi2012}.
A linear relation can still be found with fewer or more bins; gradient and intercept of the found rms-flux relation are not affected by the number of bins.

The linear relation between mean flux and rms variability is statistically quantified by the Spearman's rank correlation coefficient.
Whenever this result was close to 1, implying a positive correlation between rms and flux, a linear model was fitted to the data by using a linear regression method. 
The goodness-of-fit of the linear model is quantified by its reduced chi-squared value.
The errors on the gradient and intercept of the fitted linear model are obtained by bootstrapping the original light curve. Bootstrapping is a non-parametric resampling method used to estimate the sampling distribution underlying a statistic of interest. In this study, the statistics of interest are the gradient and intercept of the fitted linear rms-flux relation. 
A bootstrapped light curve is produced by resampling the original light curve with replacement, to which we apply the previously described data analysis method. This process is repeated 500 times. The standard deviations of the obtained sampling distributions are then adopted as the errors on their respective statistic of interest.
Note that because we have subtracted the mean flux value from the light curve, the correlation between gradient and intercept is removed. 
This is done as to produce circular error contours.

%%%%%%%%%%%%%%%%%%%%%%%%%%%%%%%%%%%%%%%%%%%%%%%%%%%%%%%%%%%%%%%%%%%%%%%%%%%%%%
\section{Results}
%%%%%%%%%%%%%%%%%%%%%%%%%%%%%%%%%%%%%%%%%%%%%%%%%%%%%%%%%%%%%%%%%%%%%%%%%%%%%%

We have analysed all CVs listed in table \ref{table:allCVs}, except for the previously studied MV Lyr.
The rms-flux relation has been detected in all steady data of the DN - SU UMa star V1404 Cyg and the NL KIC 8751494. These results will be discussed in section \ref{subsect:3-detections}. 
No rms-flux relation was detected in the other six CVs, but this can be accounted for by the low signal-to-noise (SNR) level in their light curves (see section \ref{subsect:3-others} below).

%-----------------------------------------------------------------------------
\subsection{Detection of the rms-flux relation in V1504 Cyg and KIC 8751494}	\label{subsect:3-detections}
%-----------------------------------------------------------------------------

\begin{figure}
	\center
	\begin{subfigure}[t]{0.69\columnwidth}
	\centering
	\includegraphics[width=\columnwidth]{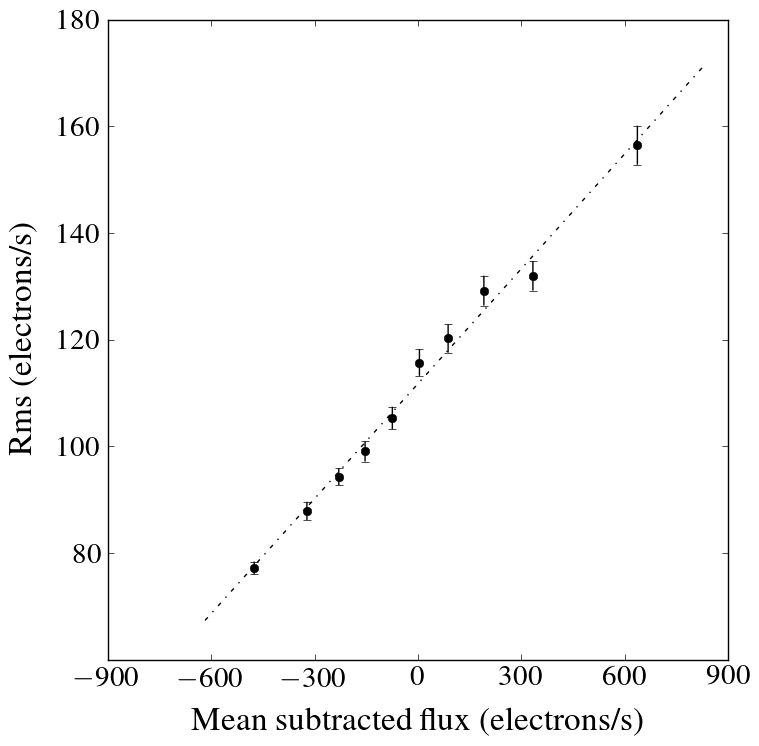}
	\captionsetup{justification=centering}
	\caption{V1504 Cyg}
	\label{subfig:Results-V1504Cyg-10dT}◊
	\end{subfigure}
	
	\begin{subfigure}[t]{0.69\columnwidth}
	\centering
	\includegraphics[width=\columnwidth]{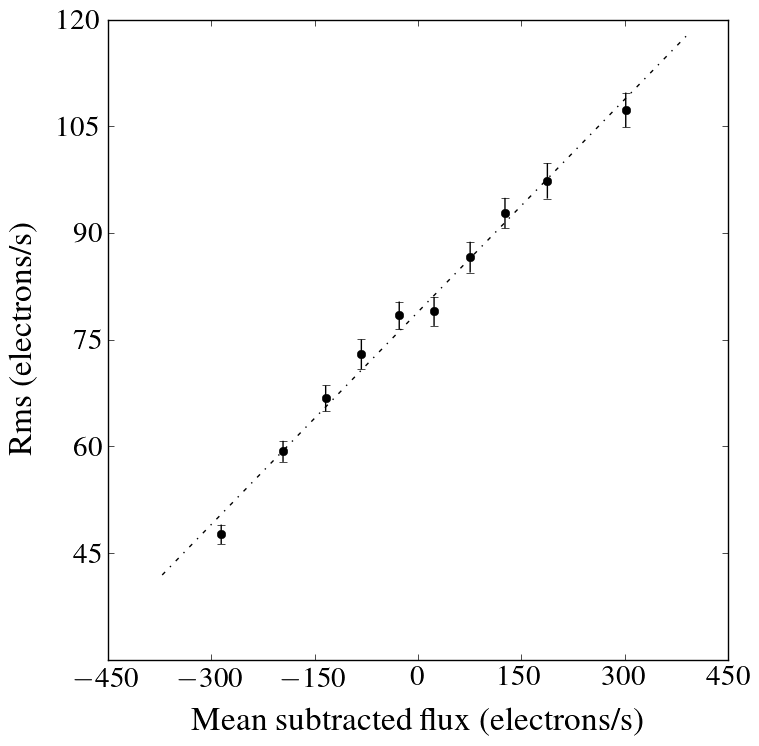}
	\captionsetup{justification=centering}
	\caption{KIC 8751494}
	\label{subfig:Results-KIC8751494-10dT}
	\end{subfigure}	
	\caption[Results combined]
	{Rms-flux relations found in the combined steady intervals of Q4 of V1504 Cyg and in the quarter Q2 of KIC 8751494. The time-scale sampled is approximately 10 minutes. 
	} 
	\label{fig:Results-RMS-10dT}
\end{figure}

The rms-flux relation has been detected in all steady intervals of V1504 Cyg, both separately and combined according to their quarter of observation, and all quarters of KIC 8751494. The total amount of steady data analysed is listed in table \ref{table:allCVs}.
Moreover, the data sets have been analysed at multiple time-scales and the rms-flux relation is found at every sampled time-scale. 
Figure \ref{fig:Results-RMS-10dT} shows the rms-flux relations found in the combined steady intervals of Q4 of V1504 Cyg and in quarter Q2 of KIC 8751494. The time-scale sampled is equal to ten times the sampling cadence, approximately ten minutes. 
The fractional rms level of the DN-SU UMa V1504 Cyg and the NL KIC 8751494 is approximately equal to 0.15 and 0.05 respectively, which is calculated by dividing the rms variability of an entire steady data set by its original mean flux. 
% Temporal evolution
{The rms-flux relation does not show any temporal evolution in both objects, unlike MV Lyr \citep{Scaringi2012}.}

In all data sets, we find that the Spearman's rank coefficient is never below 0.9. The reduced chi-squared values of linear fits shown in figure \ref{fig:Results-RMS-10dT} are 1.02 and 1.14 respectively. 
For most other data sets, we find that $\chi^2_{red} = 1 \pm \sqrt{2/K} = 1 \pm 0.5$ with $K = 8$ the degrees of freedom of our model. For some data sets in V1504 Cyg, we found $\chi^2_{red} \gg 1$. This is because the orbital period of the system is clearly visible during those intervals, which affects the high frequency aperiodic variability power.

%-----------------------------------------------------------------------------
\subsection{Conditions for the detection of the rms-flux relation}	\label{subsect:3-others}
%-----------------------------------------------------------------------------

\begin{figure*}
	\center
	\begin{subfigure}[t]{.3187\textwidth}
	\centering
	\includegraphics[width=\columnwidth]{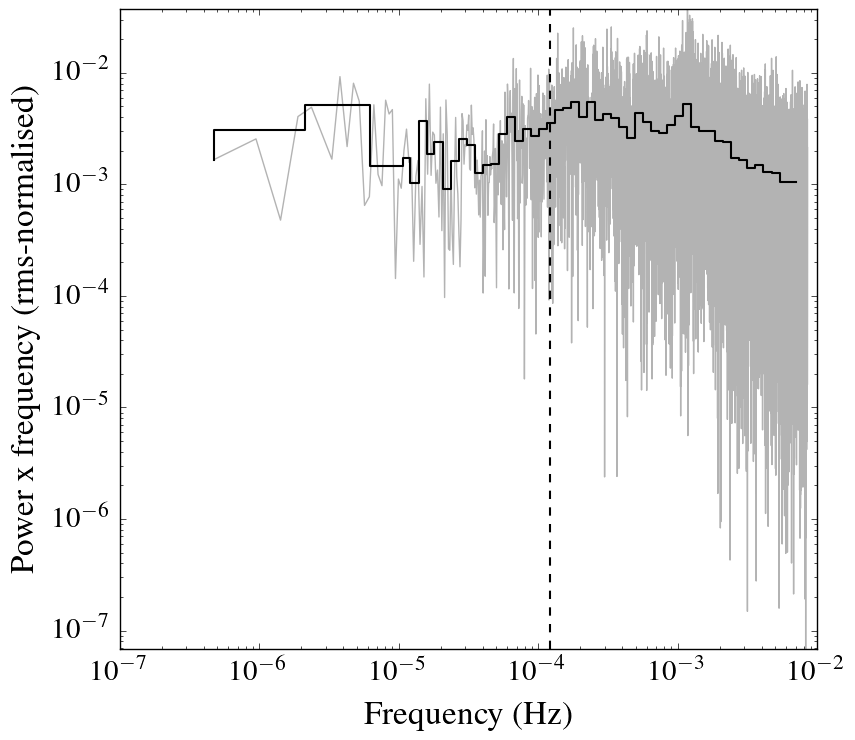}
	\captionsetup{justification=centering}
	\caption{V1504 Cyg}
	\label{subfig:Results-V1504Cyg-PSD}◊
	\end{subfigure}
	\quad
	\begin{subfigure}[t]{.3187\textwidth}
	\centering
	\includegraphics[width=\columnwidth]{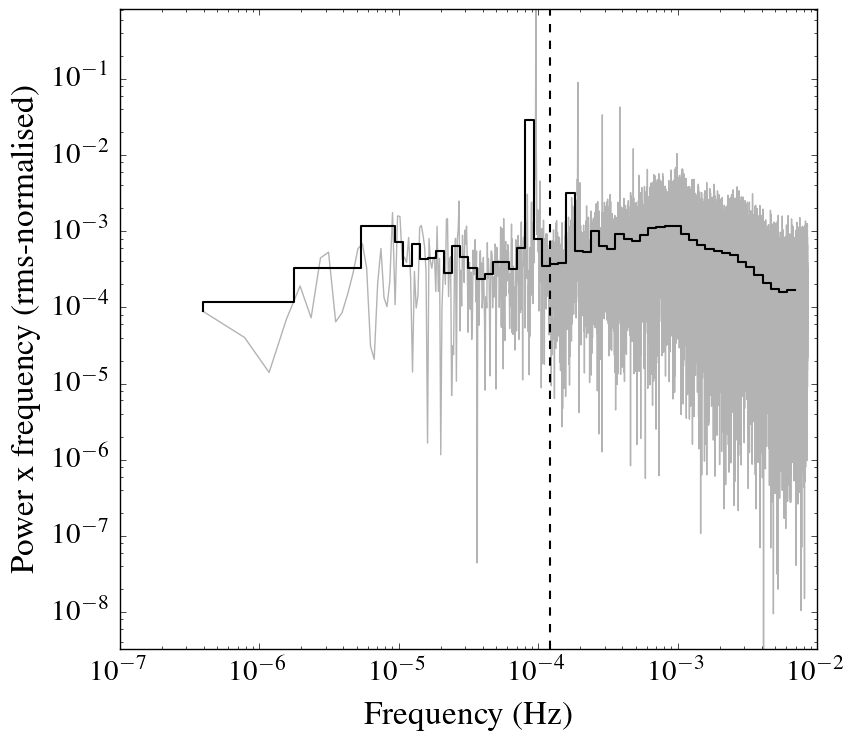}
	\captionsetup{justification=centering}
	\caption{KIC 8751494}
	\label{subfig:Results-KIC-PSD}
	\end{subfigure}	
	\quad
	\begin{subfigure}[t]{.3187\textwidth}
	\centering
	\includegraphics[width=\columnwidth]{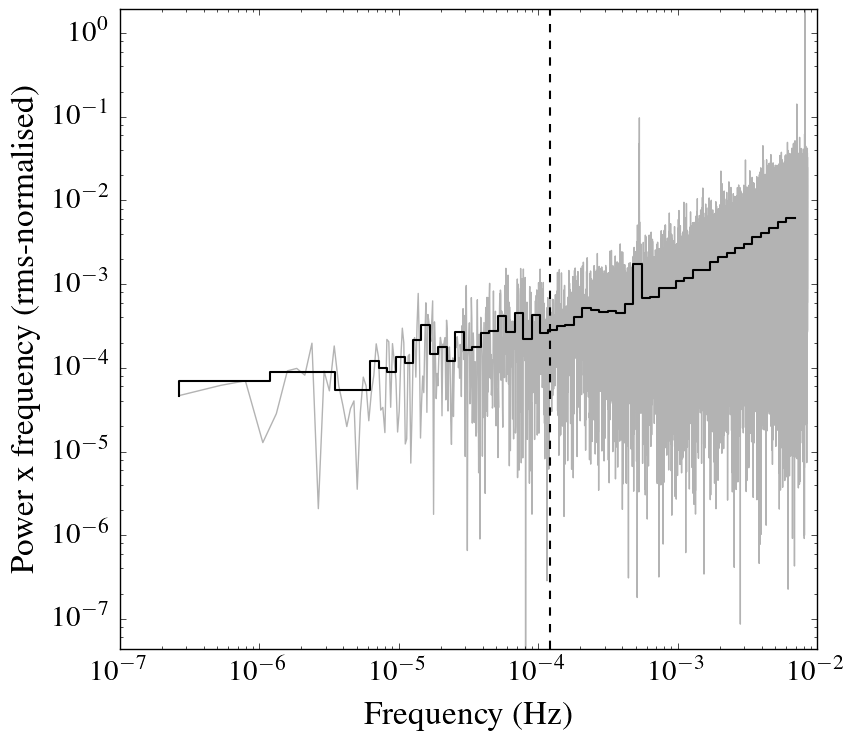}
	\captionsetup{justification=centering}
	\caption{BOKS 45906}
	\label{subfig:Results-V585-PSD}
	\end{subfigure}	
	\caption[PSDS]
	{{Representative PSDs of V1504 Cyg, KIC 8751494, and BOKS 45906 in grey. Binned PSDs are shown in black.}
	The lowest frequency sampled is marked by a dashed line, which corresponds to a time-scale of approximately 140 minutes. 
	High-frequency variability is clearly visible in the PSDs of V1504 Cyg and KIC 8751494, while Poisson noise dominates the PSD of BOKS 45906, especially at high frequencies. The PSDs of the other five CVs that do not show the rms-flux relation are similar to figure \ref{subfig:Results-V585-PSD}.
	} 
	\label{fig:Results-PSD}
\end{figure*}

\begin{table*}
\begin{minipage}{177mm}
	\centering     
	\caption{Mean flux level of a representative 10 minute segment of a steady interval for all objects, together with its rms variability and fractional rms level. 
	V1504 Cyg and KIC 8751494 are the only two systems {analysed in this paper} with detected rms-flux relations, and are also the only two systems where intrinsic variability dominates over instrumental variability. For these systems the fractional rms column shows the intrinsic value for these systems. The remaining objects are dominated by instrumental noise, and the fractional rms column is indicative on the upper limit for intrinsic variability. 
	{For completeness, MV Lyr has been included as well. {The \emph{Kepler} data of MV Lyr displayed a greater range of variability amplitudes than the objects studied in the present paper, so the values obtained for this source are expressed as a range of values \citep{Scaringi2013a}.} In this object intrinsic variability also dominates over instrumental variability.}
	} 
	\begin{tabular}{@{}l c c c c@{}}   
	\hline
	Object		&	Mean flux		& 	Rms			& 	Fractional	& {Rms dominated by} \\
				& 	(electrons)		&	(electrons)	&	rms 	   	& {intrinsic variability?}	\\
	\hline
	BOKS 45906	&   4,643    		&   645    		&   14$\%$    	& {No}	\\
	V344 Lyr  	&   26,762    		&   871    		&   3$\%$   	& {No}	\\
	V363 Lyr  	&   23,750   		&   876   	 	&   4$\%$    	& {No}	\\
	V447 Lyr  	&   2,374    		&   529    		&   22$\%$    	& {No}	\\
	V516 Lyr  	&   1,038     		&   331    		&   32$\%$   	& {No}	\\
	V585 Lyr  	&   13,238    		&   605    		&   5$\%$   	& {No}	\\
	V1504 Cyg	&   116,602  		& 	5,078 		&   4$\%$     	& {Yes}	\\
	KIC 8751494	&   242,974  		&   3,390  		&   1$\%$    	& {Yes}	\\
	{MV Lyr}		&	{3-13$\times$10$^4$}		&	{0.1-0.4$\times$10$^4$}		&	{2-5$\%$}	& {Yes}	\\
	\hline
	\end{tabular}
	\label{table:poissonrms}    
\end{minipage}
\end{table*}

Out of the eight CVs we have analysed, we have only detected the rms-flux relation in V1504 Cyg and KIC 8751494. If the rms-flux relation were truly universal, we would expect to detect it in all CVs, provided the data quality is sufficient. Here we verified whether the six systems with no detection are affected by larger noise levels, thus diminishing the SNR. Aperiodic variability cannot be detected when the instrumental noise level of the data set is comparable to or larger than its intrinsic rms variability, as the instrumental errors then `overpower' the intrinsic variability emitted by the system. 

It has already been shown by \cite{Edelson} that the instrumental noise levels within \textit{Kepler} light curves can be substantially larger than pure Poisson noise. This can possibly be attributed to Moire pattern drift (MPD), which arises from crosstalk between the four guiding sensors and the 84 science channel readouts \citep{kolo}. It has been also pointed out \citep{wehrle,revalski} that MPD noise can be a serious problem for aperiodic variability studies, especially at high frequencies for low luminosity sources. As there is no known procedure to accurately estimate this noise component we are unable to place robust constraints on the level of instrumental noise for the observed systems. 
{We have checked which objects were more heavily affected by MPD and found that only KIC 8751494 and V1504 Cyg were affected during Q5 and Q2, 6, 10, and 14 respectively \citep{Kepler}. 
The rms-flux relation is detected in all non MPD-affected, as well as MPD-affected quarters.}
However, we can inspect the PSDs of all objects to determine at which frequencies instrumental noise begins to dominate the intrinsic aperiodic variability. This is because instrumental noise will appear as white noise in the PSDs (constant power across all frequencies), whilst accreting systems are known to display strictly negative power-law indices (with breaks at specific characteristic frequencies).

{In figure {\ref{fig:Results-PSD}}, PSDs calculated with a Lomb-Scargle method for V1504 Cyg, KIC 8751494, and BOKS 45906 are shown in grey. Overplotted in black are the smoothed PSDs, calculated by binning the PSD and averaging over the bins. We have opted for a Lomb-Scargle method as the data sets are unevenly spaced due to the presence of temporal gaps.}
The lowest frequency we have sampled when searching for the rms-flux relation is marked by a vertical dashed line. From this figure it can be seen that the PSD of BOKS 45906 displays a white noise dominated PSD at frequencies above $10^{-5}$Hz (with positive power law index in power x frequency). Thus in this system any intrinsic variability cannot be measured above that frequency. The PSD shape of BOKS 45906 is representative of the PSDs of other objects which do not show the rms-flux relation in our sample, as they are also dominated by instrumental white noise. We therefore conclude that the rms-flux relation is detected within all objects for which the data quality is sufficient for its detection.

We can however try and place upper limits on the level of intrinsic rms for systems where the rms-flux relation is not detected. Table \ref{table:poissonrms} shows the mean flux and rms levels for representative 10-minute segments (the fastest timescale probed in this work), together with the measured fractional rms. These values are indicative for the intrinsic variability for V1504 Cyg and KIC 8751494, which is between $1\%$ and $4\%$, as they are the objects where intrinsic variability dominates over instrumental noise. For the other six systems the fractional rms is an upper limit on the intrinsic variability. For example, the limit obtained for BOKS 45906 is $14\%$ fractional rms variability. If this system had to display an rms-flux relation on the $1\%-4\%$ level as in KIC 8751494 and V1504 Cyg respectively, we would have not been able to detect it with the current data. The same is true for all other non-detections. In the cases of V344 Lyr, V363 Lyr and V585 Lyr, the upper limits obtained are close to the expected intrinsic rms. However, we note that in order to detect an rms-flux relation, the intrinsic rms variability of the system has to be a multiple of the instrumental rms. Because of this the quoted upper limits are somewhat conservative.

%%%%%%%%%%%%%%%%%%%%%%%%%%%%%%%%%%%%%%%%%%%%%%%%%%%%%%%%%%%%%%%%%%%%%%%%%%%%%%
\section{Discussion}
%%%%%%%%%%%%%%%%%%%%%%%%%%%%%%%%%%%%%%%%%%%%%%%%%%%%%%%%%%%%%%%%%%%%%%%%%%%%%%

\begin{figure}
	\begin{subfigure}{0.235\textwidth}
	\centering
	\includegraphics[width=\columnwidth]{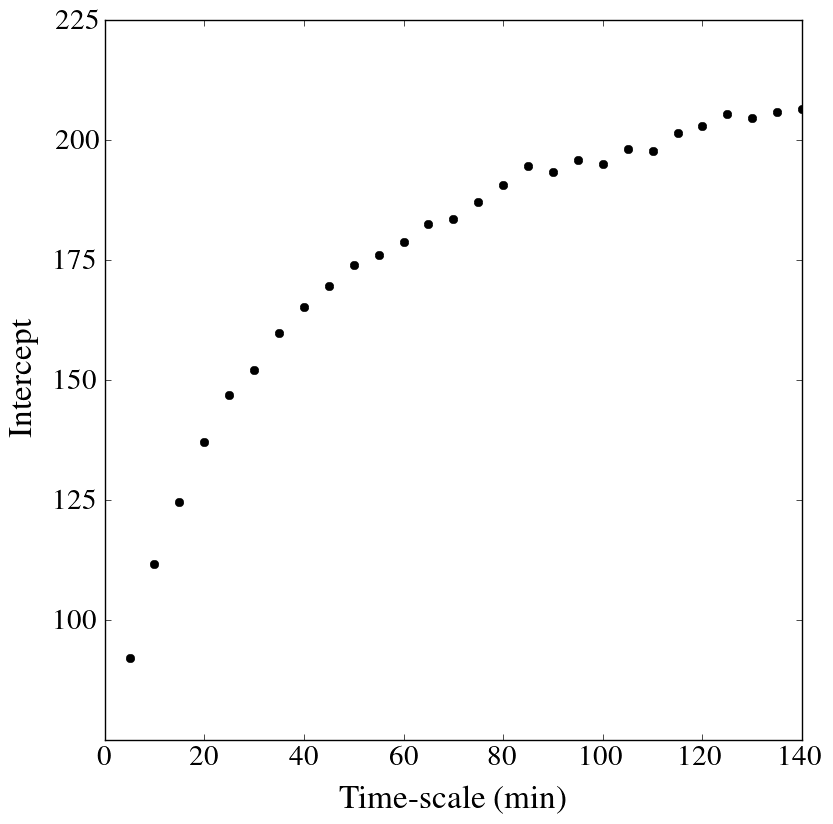}
	\captionsetup{justification=centering}
	\caption{Intercept}
	\label{subfig:Discussion-Timescale-V1504Cyg-Intercept}◊
	\end{subfigure}
	\quad
	\begin{subfigure}{0.235\textwidth}
	\centering
	\includegraphics[width=\columnwidth]{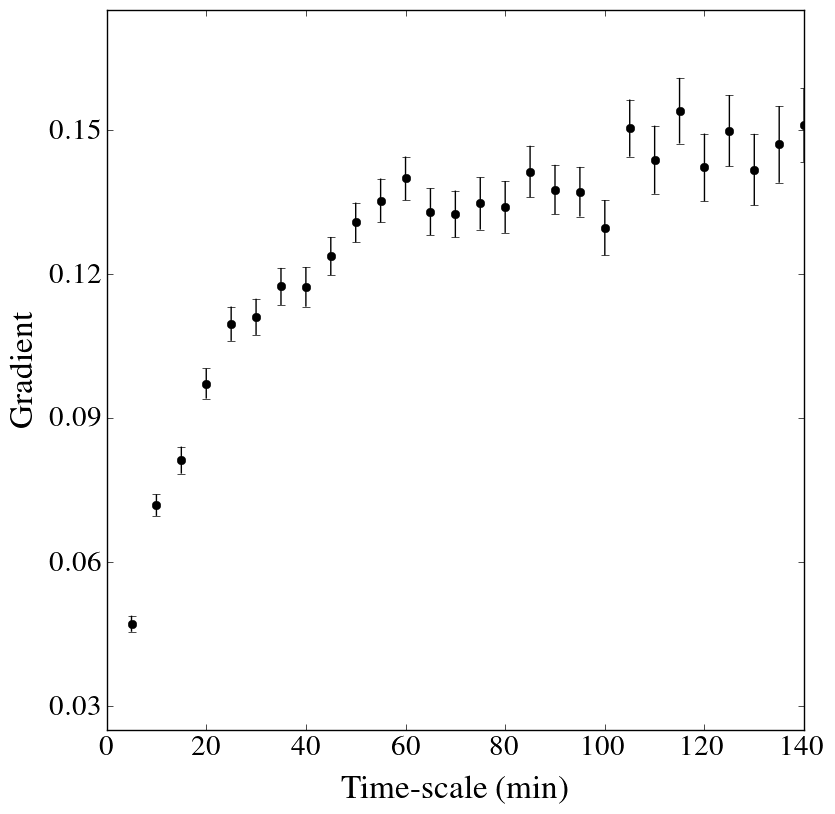}
	\captionsetup{justification=centering}
	\caption{Gradient}
	\label{subfig:Discussion-Timescale-V1504Cyg-Gradient}◊
	\end{subfigure}
	\caption[Results combined]
	{Variation of the intercept and gradient of the fitted rms-flux relation with time-scale in the combined quiescent intervals of Q4 of V1504 Cyg. The errors on the intercepts are too small to be visible; the average error is equal to 0.2 electrons/s.} 
	\label{fig:Discussion-Timescale-V1504Cyg}
\end{figure}

\begin{figure}
	\begin{subfigure}{0.235\textwidth}
	\centering
	\includegraphics[width=\columnwidth]{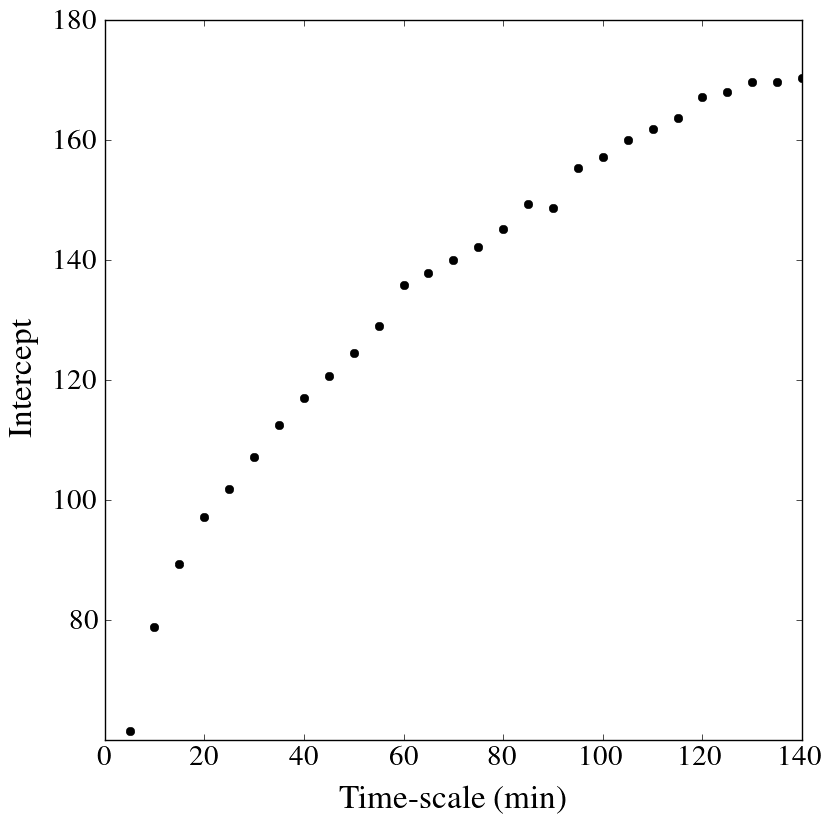}
	\captionsetup{justification=centering}
	\caption{Intercept}
	\label{subfig:Discussion-Timescale-KIC8751494-Intercept}
	\end{subfigure}	
	\quad
	\begin{subfigure}{0.235\textwidth}
	\centering
	\includegraphics[width=\columnwidth]{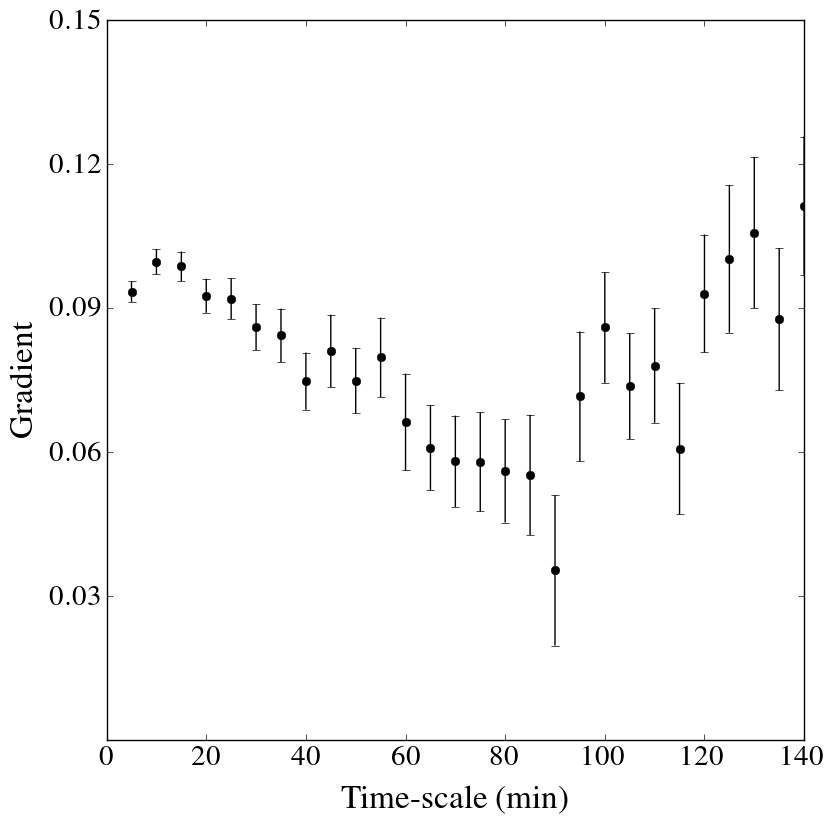}
	\captionsetup{justification=centering}
	\caption{Gradient}
	\label{subfig:Discussion-Timescale-KIC8751494-Gradient}
	\end{subfigure}	
	\caption[Results combined]
	{Variation of the intercept and gradient of the fitted rms-flux relation with time-scale in the quarter Q2 of KIC 8751494. The errors on the intercepts are too small to be visible; the average error is equal to 0.08 electrons/s.} 
	\label{fig:Discussion-Timescale-KIC8751494}
\end{figure}

The rms-flux relation is detected in the quiescent states of the DN-SU UMa V1504 Cyg and the NL KIC 8751494. 
Since the previously studied MV Lyr is also a NL, its detection in V1504 Cyg is the first detection of the rms-flux relation in a dwarf nova. 
For the other six CVs listed in table \ref{table:allCVs}, the data was not of sufficient SNR to detect the rms-flux relation.

We found that the parameters of the fitted rms-flux relation, i.e. gradient and intercept, vary as a function of the time-scale sampled during the analysis. This is observed in both V1504 Cyg and KIC 8751494. The variations however depend on the object.
We will illustrate them by analysing the combined steady intervals of Q4 for V1504 Cyg, and quarter Q2 for KIC 8751494. By analysing data from individual quarters, we ensure that we are not biased towards the inter-quarter flux offsets.
The time-scales sampled range from approximately 5 to 140 minutes. The variations seen in figures \ref{fig:Discussion-Timescale-V1504Cyg} and \ref{fig:Discussion-Timescale-KIC8751494} are representative of those found in all other data sets for both objects.

Figures \ref{subfig:Discussion-Timescale-V1504Cyg-Intercept} and \ref{subfig:Discussion-Timescale-KIC8751494-Intercept} show the variation of the intercept with time-scale.
The variation is similar for both objects: the intercept increases with time-scale. This increase flattens at longer time-scales and steepens at shorter time-scales.
The intercept of the rms-flux relation is a measure of the constant variability power observed.
%, as it corresponds to the rms variability at zero flux. 
The flattening at longer time-scales hence corresponds to the PSD flattening at frequencies below the local viscous frequencies in the disc, which are visible in the PSD as a `kink'.

Figures \ref{subfig:Discussion-Timescale-V1504Cyg-Gradient} and \ref{subfig:Discussion-Timescale-KIC8751494-Gradient} show the variation of the gradient with time-scale. 
The variation observed depends on the object: in V1504 Cyg the gradient increases with increasing time-scale, while in KIC 8751494 the gradient decreases until approximately 80 minutes, after which it increases.
The gradient of the rms-flux relation is a measure of how fast the total variability power increases with increasing flux. Its variation hence cannot be explained as straightforwardly as the variation of the intercept. 
The variation is different for both objects, which might point towards different properties of their respective accretion discs. This could potentially be attributed to the fact that V1504 Cyg is a SU UMa-type DN and KIC 8751494 a NL.

We have attempted to check whether gradient and intercept also vary over the full data set (over all steady intervals and/or quarters). However, this analysis is not straightforward to perform due to the inter-quarter flux offsets. 
We did subtract the mean flux level before analysing a steady interval, but are unsure whether this fully accounts for the inter-quarter flux offsets.
Although there seem to be changes in both parameters over the full observations, we have not quantified these variations. 
{A simple subtraction of the mean flux level is likely sufficient, given the relatively constant flux level during steady intervals.}
This will be addressed in more detail in future.

\begin{figure}
	\center
	\begin{subfigure}[t]{0.6\columnwidth}
	\centering
	\includegraphics[width=\columnwidth]{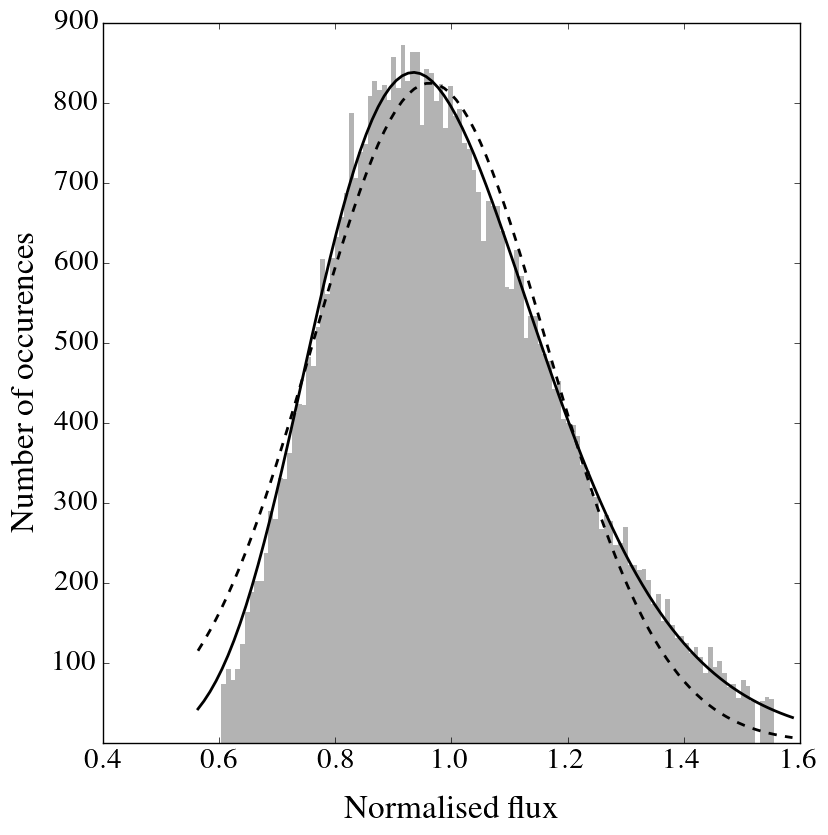}
	\captionsetup{justification=centering}
	\caption{V1504 Cyg}
	\label{subfig:Discussion-Distr-V1504Cyg}◊
	\end{subfigure}
	
	\begin{subfigure}[t]{0.6\columnwidth}
	\centering
	\includegraphics[width=\columnwidth]{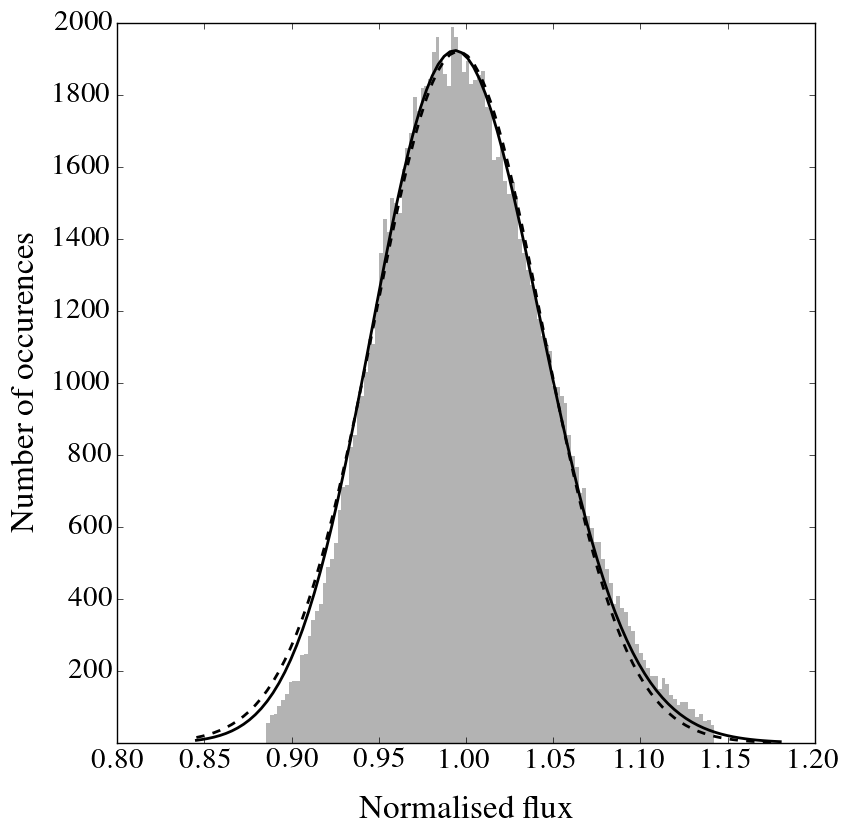}
	\captionsetup{justification=centering}
	\caption{KIC 8751494}
	\label{subfig:Discussion-Distr-KIC}
	\end{subfigure}	
	\caption[Results combined]
	{{Flux distribution obtained from the combined steady intervals of Q4 for V1504 Cyg and Q16 of KIC 8751494. The flux has been normalised by dividing through the mean flux level of the entire data set. Lognormal distribution fits are shown as a solid black line, Gaussian distribution fits are shown as a dashed black line. }}
	\label{fig:Discussion-Distr}
\end{figure}

{Another characteristic of aperiodic variability is a lognormal flux distribution \citep{Uttley2005}.
In order to check whether V1504 Cyg and KIC 8751494 exhibit a lognormal distribution as well, we have performed both a Gaussian and lognormal fit to their flux distributions. The flux distribution is obtained by producing a histogram of the fluxes and removing all bins with less than 50 measurements, similar to \citet{Scaringi2012}. Again, we have measured the goodness-of-fit by means of its reduced chi-squared value.
The results are shown in figure {\ref{fig:Discussion-Distr}}, where we have fitted the flux distributions of the combined steady intervals of Q4 for V1504 Cyg and Q16 of KIC 8751494.
The distributions clearly appear lognormal and this is also reflected in the reduced chi-squared value of the respective fits. For V1504 Cyg, the Gaussian fit resulted in a reduced chi-squared value of 14, while the lognormal fit resulted in one of 2. For KIC 8751494, the Gaussian fit resulted in a reduced chi-squared value of 15, while the lognormal fit resulted in one of 6.5. These results are representative of the flux distributions found in their other data sets. 
The large deviation from one of the reduced chi-squared value of the lognormal fit in KIC 8751494 might be due to the system's orbital period{, which might add a non-stochastic source of variability into the light curve.}
The lognormal distribution is therefore more suitable to fit the flux distributions of V1504 Cyg and KIC 8751494.}

%%%%%%%%%%%%%%%%%%%%%%%%%%%%%%%%%%%%%%%%%%%%%%%%%%%%%%%%%%%%%%%%%%%%%%%%%%%%%%
\section{Conclusion}
%%%%%%%%%%%%%%%%%%%%%%%%%%%%%%%%%%%%%%%%%%%%%%%%%%%%%%%%%%%%%%%%%%%%%%%%%%%%%%

We have probed the high-frequency variability of eight CVs in the \emph{Kepler} field-of-view. Together with the previously studied MV Lyr, these systems are all known CVs for which short cadence \emph{Kepler} data is available.
Only steady data was used for the data analysis as outbursts dominate the light curve, making it difficult to detect aperiodic variability. The steady intervals have been visually selected using a conservative range, ensuring that the system is fully in its steady state. These steady intervals have also been combined according to their quarter of observation. For systems that did not display outbursts, the quarters have been analysed as a whole.

The linear rms-flux relation is detected in two out of eight CVs: DN-SU UMa V1504 Cyg and NL KIC 8751494. Moreover, we have detected the rms-flux relation in every steady data set of both objects. 
It is the first time the rms-flux relation has been detected in a DN, as both MV Lyr and KIC 8751494 are NLs.
Because of the analogy between low-hard and high-soft state XRBs and quiescent DNe and NLs respectively, it is also the first time the rms-flux relation has been detected in a low-hard XRB analogue. 
The fractional rms values for the two systems also correspond to their analogous XRB states: V1504 Cyg is characterised by a higher fractional rms when compared to KIC 8751494.
We have determined that the data quality for the remaining six systems is not sufficient to detect the relation. The linear rms-flux relation is hence found within all systems for which the data has a high enough SNR to not be dominated by instrumental noise at high frequencies.

When analysing the rms-flux relation in V1504 Cyg and KIC 8751494, we found that the gradient and intercept of the fitted rms-flux relation found in a steady data set vary with the time-scale sampled. We attribute the variation of the intercept is mainly caused by the PSD shape of the data set, as the intercept is a measure of the constant variability power observed. 
The variation of the gradient cannot be explained in a similar manner. Moreover, the variation observed in V1504 Cyg is different from that observed in KIC 8751494. This might point towards different properties of their respective accretion discs, which is possibly due to the fact that V1504 Cyg is a SU UMa-type DN and KIC 8751494 a NL.
{However, the difference in gradients at different time scales cannot yet be attributed to any particular accretion disc or binary system parameters.}

The total number of CVs with a detected rms-flux relation is now three: V1504 Cyg, KIC 8751494, and the previously studied MV Lyr. It is now discovered in NLs and a DN, which are analogous to high-soft and low-hard states XRBs respectively. 
The rms-flux relation is hence not only a characteristic of aperiodic variability in XRBs and AGN: it is a ubiquitous characteristic of the aperiodic variability detected in all compact accreting systems. 
The time-scales of aperiodic variability are therefore coupled in CVs as well, implying that the physics of disc accretion is similar in CVs and in XRBs and AGN. 
As the coupling of all time-scales is intrinsic to the fluctuating accretion disc model, it is best suited for modelling accretion discs around the compact objects in CBs.

%%%%%%%%%%%%%%%%%%%%%%%%%%%%%%%%%%%%%%%%%%%%%%%%%%%%%%%%%%%%%%%%%%%%%%%%%%%%%%
\section*{Acknowledgements}
%%%%%%%%%%%%%%%%%%%%%%%%%%%%%%%%%%%%%%%%%%%%%%%%%%%%%%%%%%%%%%%%%%%%%%%%%%%%%%

This research has made use of the SIMBAD database, operated at CDS, Strasbourg, France.
This paper includes data collected by the \emph{Kepler} mission. Funding for the \emph{Kepler} mission is provided by the NASA Science Mission directorate.
All of the data presented in this paper were obtained from the Mikulski Archive for Space Telescopes (MAST). STScI is operated by the Association of Universities for Research in Astronomy, Inc., under NASA contract NAS5-26555. Support for MAST for non-HST data is provided by the NASA Office of Space Science via grant NNX13AC07G and by other grants and contracts. 
This research has made use of the Python repository of the Instituut voor Sterrenkunde (KU Leuven).
S.S. acknowledges funding from the FWO Pegasus Marie Curie Fellowship program. S.S. also acknowledges the Max-Planck-Institute for Extraterrestrial Physics for visiting.
{We acknowledge the useful referee report from the anonymous referee.}

\bibliographystyle{mn2e}
\bibliography{allreferences}

\label{lastpage}

\end{document}